\documentclass[11pt,twoside]{article}

\usepackage{asp2004}
\usepackage{epsf}
\usepackage{psfig}
\usepackage{lscape}

\begin{document}
\title{The fraction of DA white dwarfs with kilo-Gauss magnetic fields
}   
\author{S. Jordan,$^1$ R. Aznar Cuadrado,$^2$ R. Napiwotzki.$^3$
H.~M.~Schmid$^4$, S.~K.~Solanki$^2$}   
\affil{$^1$Astronomisches Rechen-Institut, Zentrum f\"ur Astronomie der Universit\"at Heidelberg, D-69120 Heidelberg, Germany\\
        $^2$Max-Planck-Institut f\"{u}r Sonnensystemforschung,
       D-37191 Katlenburg-Lindau, Germany\\
        $^3$Centre for Astrophysics Research, University of Hertfordshire, Hatfield AL10 9AB, UK\\
$^4$Institut f\"ur Astronomie, ETH Z\"urich,
       CH-8092 Z\"{u}rich, Switzerland
}    

\begin{abstract} 
 Current estimates for white dwarfs with fields in excess of 1\,MG are
   about 10\%; according to our first high-precision
   circular-polarimetric study of 12 bright white dwarfs with the VLT
   \citep{Aznar-etal:04} this number increases up to about 25\%\ in
   the kG regime.  With our new sample of 10 white dwarf observations
   (plus one sdO star) we wanted to improve the sample statistics to
   determine the incident of kG magnetic fields in white dwarfs. In
   one of our objects (LTT\,7987) we detected a statistically
   significant (97\%\ confidence level) longitudinal magnetic field
   varying between ($-1\pm 0.5$)\,kG and ($+1\pm 0.5$)\,kG.  This
   would be the weakest magnetic field ever found in a white dwarf,
   but at this level of accuracy, systematic errors cannot completely
   be ruled out.  Together with previous investigations, the fraction
   of kG magnetic fields in white dwarfs amounts to about $11-15$\% ,
   which is close to the current estimations for highly magnetic white
   dwarfs ($>$1\,MG).
\end{abstract}


\section{Introduction}   
Until recently, magnetic fields below  30 kG could not be detected with
the exception of the very bright white dwarf 40 Eri B ($V=8.5$), in which
\cite{Fabrika-etal:03} found a magnetic field of 4\,kG.
However, by using the ESO VLT, we could push the detection limit
down to about 1\,kG in our first investigation of  12  DA white dwarfs
with $11<V<14$ \citep{Aznar-etal:04}. In 3 objects of this sample
we detected magnetic fields between 2\,kG and 7\,kG on a $5\sigma$
confidence level. Therefore, we concluded that the fraction of white dwarfts
with kG magnetic fields is
about 25\%.
For one of our cases, LP\,672$-$001 ({WD\,1105$-$048}),  \cite{Valyavin-etal:06}
confirmed the presence of a kG magnetic field.

\section{Observations}
The spectropolarimetric data of our new sample of  ten bright normal
DA white dwarfs plus one high-metallicity sdO star 
were obtained in service mode between May 5 and August
4, 2004, with the FORS1 spectrograph at the 8\,m UT2 of the
VLT. 
We checked all  candidates for spectral
peculiarities and magnetic fields strong enough to be detected in intensity
spectra taken with the  high-resolution Echelle spectrograph UVES at the
Kueyen (UT2) of VLT in the course of the SPY project \citep{Napi-etal:03}.
The spectra and circular polarimetric data covered the wavelength region
between 3600\,\AA\ and 6000\,\AA\ with a spectral resolution of
4.5\,\AA.

In order to avoid saturation the exposures were split into
a sequence of exposures; after every second observation the retarder
plate was rotated from $\alpha=-45^\circ$ to $\alpha=+45^\circ$ and back
in order to suppress spurious signals in the degree of circular polarisation
(calculated from the ratio of the Stokes parameters $V$ and $I$).

\section{Determination of magnetic fields}
The theoretical $V/I$ profile
for a given mean longitudinal magnetic field $\langle B_z\rangle$
(expressed in Gauss) below
about 10\,kG is given by  the weak-field approximation
\citep[e.g.\,][]{Angel-Landstreet:70} without any loss of
accuracy:
\begin{equation}
\frac{V}{I} = -g_{\rm eff} \ensuremath{C_z}\lambda^{2}\frac{1}{I}
\frac{\partial I}{\partial \lambda}
\ensuremath{\langle\large B_z\large\rangle}\;,
\label{e:lf}
\end{equation}
where $g_{\rm eff}$ is the effective Land\'{e} factor
\citep[= 1 for all hydrogen lines of any series,][]{Casini-Landi:94},
$\lambda$ is the wavelength expressed in \AA,
and the constant
$C_z=e/(4\pi m_ec^2)$ $(\simeq 4.67 \times10^{-13}\,{\rm G}^{-1} \mbox{\AA}^{-1})$.

We performed a $\chi^2$-minimisation procedure in order to find out
which mean longitudinal magnetic field strength best fits the
observed data in wavelength  intervals  of $\pm$20\,\AA\ around H$\beta$
and H$\gamma$.
The resulting best-fit values for the magnetic field strengths from
the individual lines and their statistical $1\sigma$ errors are listed in
Table\,\ref{t:mf} for each observation. We also provided the weighted means
from both lines.

Our fitting procedure was validated with extensive numerical simulations
using a large sample of 1000 artificial noisy polarisation spectra
\citep{Aznar-etal:04}. It was concluded that at our noise level
kG fields can reliably be detected.

\section{Results}	

\begin{table*}[!ht]
\caption{Magnetic fields derived from the H$\gamma$ and H$\beta$ lines for our
sample of white dwarfs. $B({\sigma})$ provides the magnetic field in units of
the $\sigma$ level. Detections exceeding the $2\sigma$ levels are given in
bold. Multiple
observations that were averaged prior to analysis are labeled {\sc average}.}
\label{t:mf}
\begin{center}
{\footnotesize
\begin{tabular}[c]{lcr@{$\,\pm\,$}lr@{$\,\pm\,$}lr@{$\,\pm\,$}lrr}
\hline\noalign{\smallskip}
\multicolumn{1}{c}{Target} & \multicolumn{1}{c}{Date} &
 \multicolumn{4}{c}{$B$(G)} &
\multicolumn{2}{c}{$B({\rm G})$} & \multicolumn{1}{c}{$B({\sigma})$} &
\\
\multicolumn{1}{c}{} & \multicolumn{1}{c}{} &
\multicolumn{2}{c}{H$\gamma$} & \multicolumn{2}{c}{H$\beta$} &
\multicolumn{2}{c}{H$\gamma, \beta$} & \multicolumn{1}{c}{H$\gamma, \beta$} 
 \\
\noalign{\smallskip}\hline\noalign{\smallskip}
{WD\,1148$-$230}&08/05/04        &$-520$&655      &$-980$&590   &$-774$&438   & 1.76  \\
(EC\,11481$-$2303,    &18/05/04  &$-30$&1325      &20&1095      &0&844        & 0.00  \\
(sdO)            &{\sc average}  &$-490$&625      &$-860$&500   &$-716$&390   & 1.83  \\[0.2cm]
{WD\,1202$-$232}&18/05/04        &1280&865        &260&940      &812&636      & 1.28  \\
(EC\,12028$-$2316)&25/05/04      &660&550         &$-370$&325   &$-103$&280   & 0.37  \\
            &{\sc average}       &$-200$&260      &850&425      &85&221       & 0.39 \\[0.2cm]
{WD\,1327$-$083}&25/05/04        &300&1010        &$-320$&1080  &10&737       & 0.01  \\
(G\,14$-$058)  &27/05/04         &$-2800$&740     &2790&835     &$-340$&553   & 0.61  \\
            &{\sc average}       &$-1230$&555     &1410&520     &175&379      & 0.46 \\[0.2cm]
{WD\,1620$-$391}&10/05/04        &150&785         &$-1580$&475  &$\mathbf{-1116}$&{\bf 406} & 2.75   \\
(CD$-$38\,10980)    &17/05/04    &$-$2390&1220    &$-20$&735    &$-651$&629   & 1.03 \\
            &25/05/04            &120&640         &$-770$&595   &$-357$&435   & 0.82  \\
            &{\sc average}       &$-500$&480      &$-920$&365   &$\mathbf{-766}$&{\bf 290} & 2.63 \\[0.2cm]
{WD\,1845+019}&05/05/04          &$-340$&1410     &$-150$&1175  &$-227$&902   & 0.25  \\
(Lan\,18)     &10/05/04          &$-30$&1145      &6000&2390    &1095&1032    & 1.06  \\
            &{\sc average}       &$-130$&815      &500&670      &245&517      & 0.47  \\[0.2cm]
{WD\,1919+145}&06/05/04          &1240&1080       &$-1440$&1220 &62&808       & 0.08  \\
(GD\,219)     &10/05/04          &930&1290        &110&990      &413&785      & 0.53  \\
            &{\sc average}       &1180&870        &$-500$&815   &285&594      & 0.48   \\[0.2cm]
{WD\,2007$-$303}&06/05/04        &$-1460$&1270    &$-1040$&485  &$\mathbf{-1093}$&{\bf 453} & 2.41  \\
(LTT\,7987)   &12/05/04          &1540&780        &610&620      &$\mathbf{970}$&{\bf 485}  & 2.00  \\
            &{\sc average}       &120&495         &$-390$&375   &$-204$&298   & 0.67  \\[0.2cm]
{WD\,2014$-$575}&14/05/04        &2410&1735       &1240&1260    &1643&1019    & 1.61  \\
(RE\,2018$-$572)     &27/06/04   &310&4295        &1160&3340    &839&2636     & 0.32  \\
            &28/06/04            &2820&2060       &$-4470$&1455 &$-2043$&1188 & 1.71  \\
            &{\sc average}       &2380&1205       &$-1120$&895  &124&718      & 0.17  \\[0.2cm]
{WD\,2039$-$202}&17/05/04        &$-2670$&1595    &40&965       &$-686$&825   & 0.83  \\
(LTT\,8189)    &10/06/04         &$-780$&730      &$-1800$&720  &$\mathbf{-1297}$&{\bf 512} & 2.53  \\
            &{\sc average}       &$-1240$&655     &$-1290$&535  &$\mathbf{-1269}$&{\bf 414}   & 3.06  \\[0.2cm]
{WD\,2149+021}&26/06/04          &340&1060        &730&930      &560&699      & 0.80  \\
(G\,93$-$048)  &09/07/04         &$-530$&945      &$-1690$&890  &$-1144$&647  & 1.78  \\
            &04/08/04            &$-1300$&985     &350&705      &$-208$&573   & 0.36  \\
            &{\sc average}       &$-600$&555      &$-130$&490   &$-335$&367   & 0.91  \\[0.2cm]
{WD\,2211$-$495}&14/05/04        &110&1655        &$-1940$&1060 &$-1343$&892  & 1.51  \\
(RE\,2214$-$491)   &28/06/04     &$-190$&1795     &640&1190     &386&991      & 0.39  \\
           &{\sc average}        &$-390$&1155     &$-900$&795   &$-736$&654   & 1.12  \\
\noalign{\smallskip}\hline
\end{tabular}
}
\end{center}
\end{table*}

As can be seen from Table\,\ref{t:mf}
 none of the measurements of the circular polarisations
reached the same level of confidence as the three magnetic objects
found in the first sample (Paper\,I). The highest level of
confidence was achieved by {LTT\,7987} ({WD\,2007$-$303}) where a
$2.4$ and $2.0\sigma$ level was reached for the two respective observations.
The corresponding mean longitudinal field strengths were
$-1093\pm 453$\,G and $970\pm 485$\,G. Single observations  of
{CD$-$38 10980} ({WD1620$-$391}) resulted in $-1116\pm 406$\,G and
{LTT\,8189} ({WD\,2039$-$202})  in $-1297\pm 512$\,G, which
corresponds to $2.8\sigma$ and $2.5\sigma$, respectively.

With two
observations exceeding $2\sigma$, {LTT\,7987} ({WD\,2007$-$303})
would be the most convincing case for  being a positive detection.
The probability that two independent
and uncorrelated observations
of a single star have that level of confidence can be estimated
in the following way:
The likelihood that an observations exceeds 2$\sigma$ is 4.6\%. Therefore,
the chance that at least one observation  of the white dwarfs exceeds 2$\sigma$ is
$(1-0.954^{23})=66.1$\%. Then the probability that the same star has a
second observation exceeding 2$\sigma$ is $0.661\cdot 0.046=3.0$\%.
Therefore,
from a purely statistical point of view we must regard this detection
as significant (with 97\%\ confidence).

Both measurements of the sdO star EC\,11481-2303 are below the $2\sigma$ level.
This is interesting by itself and confirms the finding by \cite{otooleetal05-2} that
there is no correlation between the metallicity and the presence of a magnetic
field with kG strength.

\section{Conclusion}
While we detected magnetic fields in  3 out of 12 programme stars
in our first investigation,
we found at most (if at all) one object in our new sample
of 10 DA white dwarfs.
Putting both samples together we arrive at a fraction of
14$-$18\%\  of kG magnetism in white dwarfs; the lower value is obtained
assuming that
LTT\,7987 is not magnetic. However, if confirmed, LTT\,7987 would
have the lowest magnetic field (1\,kG) ever detected in a white dwarf.

Recently, \cite{Valyavin-etal:06} have also performed a search for
circular polarisation in white dwarfs. They confirmed our detection
\citep{Aznar-etal:04}
of a varying longitudinal magnetic field in {LP\,672$-$001} ({WD\,1105$-$048}):
they measured field strengths between $-7.9\pm 2.6$\,kG to $0.1\pm2.7$\,kG,
compared to our values of $-4.0\pm 0.7$\,kG to $-2.1\pm0.4$\,kG.
However, they did not discover any significant magnetic field in their
five other programme stars. If we combine their and our results together,
the fraction of kG magnetic fields in DA white dwarfs amounts to
15\%\ (4/(12+10+5)) or 11\%\ (3/(12+10+5)), if we disregard the detection in
LTT\,7987. However, it is  problematic to merge both samples, because
the signal-to-noise ratio of our VLT measurements is  much higher than
the observations with the 6\,m telescope of the Special Astrophysical Observatory. Since our
uncertainties are on the average 2$-$3 times smaller (partly also due to the
fact that \cite{Valyavin-etal:06} have used H$\alpha$ only)
we must put a higher statistical weight on our sample
with a fraction of 11\%\ to 15\%\ of magnetic to field-free
(i.e. below detection limit) white dwarfs.


\acknowledgements 
This paper is based on observations made with ESO Telescopes at the La Silla or Paranal Observatories under programme ID 073.D-0356.
We acknowledge the use of LTE model spectra computed
by D. Koester, Kiel.



\begin{thebibliography}{}
\bibitem[Angel \& Landstreet(1970)]{Angel-Landstreet:70}
         Angel J. R. P., Landstreet J. D. 1970, ApJ 160, L147
\bibitem[{Aznar Cuadrado et al.(2004)}]{Aznar-etal:04}
Aznar Cuadrado R., Jordan S., Napiwotzki R., Schmid H.M.,
Solanki S.K., Mathys G. 2004, A\&A 423, 1081
\bibitem[{Casini \& Landi degl'Innocenti(1994)}]{Casini-Landi:94}
         Casini R., Landi degl'Innocenti E. 1994, A\&A 291, 668
\bibitem[{Fabrika et al.(2003)}]{Fabrika-etal:03}
         Fabrika S. N., Valyavin G. G., Burlakova T. E. 2003,
         Astronomy Letters 29, 737
\bibitem[{Napiwotzki et al.(2003)}]{Napi-etal:03}
         Napiwotzki R., Christlieb N., Drechsel H., et al.\ 2003,
         ESO-Messenger 112, 25
\bibitem[{{O'Toole} {et~al.}(2005){O'Toole}, {Jordan}, {Friedrich}, \&
  {Heber}}]{otooleetal05-2}
{O'Toole}, S.~J., {Jordan}, S., {Friedrich}, S., \& {Heber}, U. 2005, A\&A,
  437, 227
\bibitem[{Valyavin et al.(2006))}]{Valyavin-etal:06}
         Valyavin G., Bagnulo S., Fabrika S., Reisenegger A., Wade G.A.,
         Han I., Monin D. 2006, ApJ 648, 559
\end{thebibliography}
\end{document}